# Weak Localization Effects On The Electron-Phonon Interaction In Disordered Metals

Yong-Jihn Kim

*Department of Physics,
University of Puerto Rico,
Mayaguez, PR 00681*

**Abstract.** Weak localization has a strong influence on both the normal and superconducting properties of metals. In particular, since weak localization leads to the decoupling of electrons and phonons, the temperature dependence of resistance decreases with increasing disorder, as manifested by Mooij's empirical rule. In addition, Testardi's universal correlation of $T_c$ and the resistance ratio follows. We show that weak localization leads to the similar correction to other physical quantities, such as electronic thermal resistivity, electronic heat capacity, and ultrasonic attenuation, which are controlled by the electron-phonon interaction.

**Keywords:** Weak localization, Electron-phonon interaction, Disordered metals.
**PACS:** 72.10.Di, 71.38.Cn, 72.15.Cz, 72.15.Rn.

## INTRODUCTION

Although weak localization has greatly deepened our understanding of the normal state of disordered metals,[1] its effect on superconductivity and the electron-phonon interaction was not well understood.[2] Recently, it has been shown that weak localization suppresses the electron-phonon interaction and superconductivity, which provides a microscopic basis for understanding this fundamental problem.[3]

Since electrons and phonons are decoupled due to weak localization, we expect that the temperature dependence of the resistivity is decreasing with increasing disorder, which is the essence of the Mooij's empirical rule.[4,5] In addition, we find the Testardi correlation between $T_c$ and the resistance ratio, as we increase the disorder in metals.[6] We stress that this correlation can probe the phonon mechanism in superconductors.[7] In fact, this method has provided definite proof of the phonon mechanism in $MgB_2$.[8] We show that the similar correction occurs in other physical quantities, such as electronic thermal resistivity, electronic heat capacity, and ultrasonic attenuation.

## WEAK LOCALIZATION EFFECT ON THE ELECTRON-PHONON INTERACTION

The following Table 1 shows the comparison of conductivity and phonon-mediated matrix element in dirty and weak localization limits for 3d.[3] Here $\ell$ and $L$ are the elastic and inelastic mean free paths, respectively. Note that the same weak localization correction term occurs in both the conductivity and the phonon-mediated interaction.

**TABLE 1.** Conductivity and phonon-mediated matrix element in dirty and weak localization limits.

| Disorder Limit | Dirty | Weak localization |
|---|---|---|
| $\ell$ | ~100Å | ~10Å |
| $\sigma$ | $\sigma_B$ | $\sigma_B \left[ 1 - \dfrac{3}{(k_F \ell)^2}(1 - \dfrac{\ell}{L}) \right]$ |
| $V_{nn'}$ | $V$ | $V \left[ 1 - \dfrac{3}{(k_F \ell)^2}(1 - \dfrac{\ell}{L}) \right]$ |

Note that $V \propto \alpha^2 (\approx \alpha_{tr}^2)$ in the strong coupling theory.

# THE MOOIJ RULE

At high temperatures, the phonon-limited resistivity, $\rho_{ph}$, becomes

$$\rho_{ph} = \frac{2\pi m k_B T}{ne^2 \hbar} \lambda_{tr}. \quad (1)$$

Since weak localization decreases $\lambda_{tr}$, the Mooij rule follows,[5] as shown in Fig. 1.

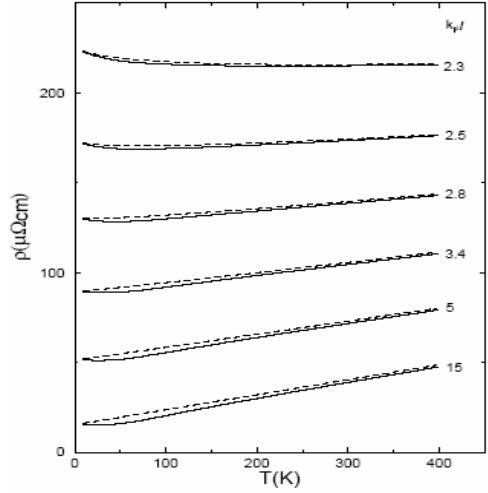

**FIGURE 1.** Calculated resistivity vs temperature with increasing disorder. The dashed lines represent resistivity obtained from Eq. (1), whereas the solid lines are from a more accurate formula (Ref. 5).

# TESTARDI CORELATION

The resistance ratio is given by

$$\frac{\rho(300K)}{\rho_0} \cong 1 + \frac{2\pi\tau \times 300K}{\hbar} \lambda_{tr}. \quad (2)$$

Since $\lambda \approx \lambda_{tr}$, Testardi correlation between $T_c$ and the resistance ratio follows, as shown in Fig. 2.[7]

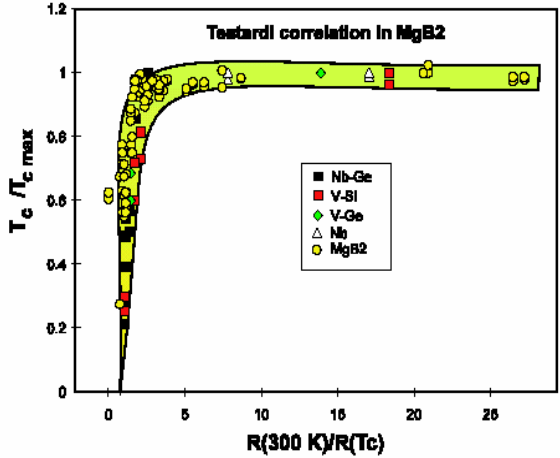

**FIGURE 2.** Testardi correlation between $T_c$ and the resistance ratio for MgB$_2$, Nb-Ge, V-Si, V-Ge, and Nb, collected by Buzea and Yamashita (Ref. 8).

# WEAK LOCALIZATION EFFECT ON THERMAL RESISTIVITY, HEAT CAPACITY, AND ULTRASONIC ATTENUATION

We expect the same weak localization correction to the electronic thermal resistivity ($1/\kappa$), the electronic heat capacity ($C_e$), and the ultrasonic attenuation. Since

$$\frac{1}{\kappa} = \frac{1}{L_0 T} \frac{4\pi m}{ne^2 k_B T} \int \frac{\hbar\omega}{(e^{\beta\hbar\omega}-1)(1-e^{-\beta\hbar\omega})}$$

$$\left\{ \left[1 - \frac{1}{2\pi^2}\left(\frac{\hbar\omega}{k_B T}\right)^2\right] \alpha_{tr}^2 F(\omega) \quad (3) \right.$$

$$\left. + \frac{3}{2\pi^2}\left(\frac{\hbar\omega}{k_B T}\right)^2 \alpha^2 F(\omega) \right\} d\omega,$$

$$C_e = \gamma T = \gamma_0(1+\lambda)T, \quad (4)$$

these quantities will decrease due to weak localization. We anticipate the similar behavior for the ultrasonic attenuation coefficient of a longitudinal ($\alpha_L$) and transverse ($\alpha_T$) wave.